\documentclass[preprint,prl,aps,onecolumn]{revtex4}
\usepackage[latin9]{inputenc}
\usepackage{mathtools}
\usepackage{amsbsy}
\usepackage{amstext}
\usepackage{graphicx}
\usepackage{braket}
\usepackage[unicode=true,pdfusetitle, bookmarks=false, breaklinks=false,pdfborder={0 0 1},backref=false,colorlinks=false]{hyperref}
\hypersetup{bookmarksnumbered=false,bookmarksopen=false}

\makeatletter

\@ifundefined{textcolor}{}
{%
 \definecolor{BLACK}{gray}{0}
 \definecolor{WHITE}{gray}{1}
 \definecolor{RED}{rgb}{1,0,0}
 \definecolor{GREEN}{rgb}{0,1,0}
 \definecolor{BLUE}{rgb}{0,0,1}
 \definecolor{CYAN}{cmyk}{1,0,0,0}
 \definecolor{MAGENTA}{cmyk}{0,1,0,0}
 \definecolor{YELLOW}{cmyk}{0,0,1,0}
}

\usepackage{amsmath}
\usepackage{graphicx}
\usepackage{amssymb}
\usepackage{txfonts,color}
\makeatother
\date{\today}

\begin{document}
\title{Lattice gauge fields via modulation in circuit QED: \\ The bosonic Creutz ladder}

\author{Hadiseh Alaeian}
\affiliation{5. Physikalisches Institut, Universitat Stuttgart, Pfaffenwaldring 57, 70569 Stuttgart, Germany}

\author{Chung Wai Sandbo Chang}
\affiliation{IQC and Electrical and Computer Engineering department, University of Waterloo, 200 University Ave. West Waterloo, Ontario, Canada}

\author{Mehran Vahdani Moghaddam}
\affiliation{IQC and Electrical and Computer Engineering department, University of Waterloo, 200 University Ave. West Waterloo, Ontario, Canada}

\author{Enrique Solano}
\affiliation{Department of Physical Chemistry, University of the Basque Country UPV/EHU, Apartado 644, 48080 Bilbao, Spain}
\affiliation{IKERBASQUE, Basque Foundation for Science, Maria Diaz de Haro 3, E-48013 Bilbao, Spain}
\affiliation{Department of Physics, Shanghai University, 200444 Shanghai, China}

\author{Christopher M. Wilson}
\affiliation{IQC and Electrical and Computer Engineering department, University of Waterloo, 200 University Ave. West Waterloo, Ontario, Canada}

\author{Enrique Rico}
\affiliation{Department of Physical Chemistry, University of the Basque Country UPV/EHU, Apartado 644, 48080 Bilbao, Spain}
\affiliation{IKERBASQUE, Basque Foundation for Science, Maria Diaz de Haro 3, E-48013 Bilbao, Spain}

\begin{abstract}
In this work we propose two protocols to make an effective gauge potential for microwave photons in circuit QED. The schemes consist of coupled transmons whose flux are harmonically modulated in time. We investigate the effect of various types of capacitive and inductive couplings, and the role of the fixed phase offset of each site on the complex coupling rate between coupled qubits. These configurations can be directly realised in a superconducting circuit and is easily extendable to a scalable lattice. Due to the intrinsic non-linearity of the transmon qubits such lattices would be an ideal platform for simulating Bose-Hubbard Hamiltonians with non-trivial gauge fields.

\end{abstract}

\maketitle
\section{Introduction}
Coherence and coherent effects are the hallmarks of quantum systems. The flourishing and growing field of circuit quantum electrodynamics (cQED) in recent years have opened a horizon in quantum control and coherent studies via benefiting from the quantized electromagnetic fields of a circuit mimicking an atom with discrete states. However, the controllability and ease of tunability of cQED elements make them powerful candidates for some of the large scale quantum networks and integration. So far, cQED are almost the only engineerable quantum system whose properties arise from the quantised electric charge and magnetic flux to make a harmonic ladder and, the Josephson junction is the main non-linear element leading to an anharmonic ladder with unequal energy spacing to make an artificial atoms. 

Photons seems to be one of the best information carriers due to their ease of control and preparation. Moreover, the recent advancements in the realm of photonics have made it possible to manipulate and steer them almost arbitrarily. Therefore, for making a large scale network of qubits, photons are one of the obvious choices. However, photons are neutral particles without any charge hence they do not lend themselves to the magnetic field manipulations, trivially. Recently an artificial gauge field has be synthesised from the atom-light interaction that controls the flow of the neutral photons as if they are charged particles and move in a magnetic field~\cite{jaksch2003creation,Dalibard11}.

In circuit QED, the atom-light interactions are implemented via a combination of the microwave resonators and superconducting qubits on an integrated chip with diverse experimental control~\cite{Devoret2004}. The system can easily be extended to a lattice scheme to realizes efficient simulators for Bose-Hubbard, and Jaynes-Cummings-Hubbard models. Moreover, due to the inherent openness of the system, cQED provides a unique platform to investigate driven-dissipative systems and study the strong correlations and non-equilibrium physics~\cite{koch2010time,nunnenkamp2011synthetic,houck2012chip,Schmidt13,PhysRevA.87.062336,Barends:2016fe,Roushan:2016fy}. Another unique feature of cQED is the possibility of studying quantum physics phenomena on a macroscopic scale. Benefiting from the inherent non-linearity of the Josephson junction the interaction between qubits can be realized, as well. 

In this article we report on the control of dynamical coupling between two superconducting qubits in a microwave circuit. We employ light-matter coupling to induce photon-photon interaction and generate an effective magnetic field for the photons. By periodically modulating the flux of a Josephson junction we will show the emergence of an effective magnetic field as a non-trivial phase in the hopping term between the adjacent sites~\cite{umucalilar2011artificial,Umucalilar12}. In contrast to other systems, the parameters of this setup like site potential, on-site interaction, and the driving frequency can be varied over a wide range. The effect of this phase on the photon transport is demonstrated in a lattice of coupled qubits in a plaquette, showing a directional photon transport along the plaquette edges. This topological feature can be further utilised to make a protected photon transport in a disordered lattice as previously proposed and demonstrated in other coupled resonator systems~\cite{Hafezi13,Lin14}.

This work is organised as follows. In the second section we briefly review the circuit QED and transmons as one of its main building blocks. After deriving the effective Hamiltonian of this system we propose a method to generate non-reciprocal hopping in a circuit of two coupled transmons. The implications of having such a complex coupling between two lattice nodes will be considered in the third section via studying the dynamics of Creutz ladder. In the fourth section we focus on one plaquette of the Creutz ladder that could be realised with the current circuit QED technology. There we explicitly show  how some of the non-trivial features of the ladder including the chiral modes and edge modes can be observed in this 4-site lattice. The study and the results present a roadmap for simulating the Bosonic Creutz ladder in a circuit QED setup. Finally, section five concludes the paper and presents some of the immediate theoretical and experimental follow up works.

\section{Circuit QED implementation}

\subsection{Transmon Qubit}

A transmon is one of the basic elements of the circuit QED described as a nonlinear LC-circuit. It consists of a capacitor in parallel with a Josephson junction. Due to the inherent nonlinearity of the latter element, the energy-ladder harmonicity of the  LC-harmonic oscillator is removed and an effective 2-level atom is obtained. The following expression gives the Hamiltonian of a transmon consists of a capacitor $C$ and a Josephson junction with energy $E_J$.

\begin{equation}
\hat{\mathcal{H}}_{\rm NLC} = \frac{\hat{Q}^2}{2C}- E_J  \cos\left( \frac{\hat{\phi}}{\phi_0}\right),
\end{equation}
where $\hat{Q}$ and $\hat{\phi}$ are charge and flux operators satisfying the canonical commutation relation of $[\hat{\phi},\hat{Q}]=i\hbar$. For small flux fluctuations we can expand the nonlinear potential $\cos(\hat{\phi}/\phi_0)$ and re-write the Hamiltonian as $\hat{\mathcal{H}}_{\rm NLC}=\hat{\mathcal{H}}_{0}+\hat{\mathcal{H}}_{1}$ as follows.

\begin{subequations}
\begin{align}
\hat{\mathcal{H}}_{0}=   \frac{\hat{Q}^2}{2C}+ \frac{E_J}{\phi_0^2} \frac{\hat{\phi}^2}{2} = \frac{\hat{Q}^2}{2C}+ \frac{\hat{\phi}^2}{2L}\\ 
\hat{\mathcal{H}}_1= - E_J  \left[ \cos\left( \frac{\hat{\phi}}{\phi_0}\right)-1+\frac{\hat{\phi}^2}{2\phi_0^2}\right].
\end{align}
\end{subequations} 

In the above equations $\hat{\mathcal{H}}_{0}$ is the Hamiltonian of a harmonic oscillator with capacitance $C$ a total inductance $L =\phi_0^2/E_J$. The remaining operator, $\hat{\mathcal{H}}_{1}$, is the nonlinear part of the Hamiltonian.

We define the normalised, dimensionless charge $\hat{q}$ and flux $\hat{\varphi}$ as

\begin{equation}\label{normalized variables}
\hat{q} = \frac{\hat{Q}}{\sqrt[4]{\hbar^2C/L}} ,~ \hat{\varphi} = \frac{\hat{\phi}}{\sqrt[4]{\hbar^2L/C}} 
\end{equation} 

Substituting these normalised variables in the harmonic oscillator Hamiltonian, the linear part $\hat{\mathcal{H}}_{0}$ could be rewritten in the canonical form of a quantum harmonic oscillator as
 
\begin{equation}
\hat{\mathcal{H}}_0= \frac{\hbar \omega_0}{2} \left( \hat{q}^2 + \hat{\varphi}^2\right) = \hbar\omega_0 \left(\hat{a}^\dag \hat{a} + \frac{1}{2}\right)
\end{equation}
where $\omega_0=  \sqrt{1/LC}$ is the resonance frequency of the linear LC-circuit, and the bosonic operators $\hat{a}$ and $\hat{a}^\dag$ are the usual annihilation and creation operators defined as $\hat{\varphi}=(\hat{a}+\hat{a}^\dag)/\sqrt{2}$ and $\hat{q}=i(\hat{a}^\dag- \hat{a})/\sqrt{2}$.

In a typical transmon qubit $E_J  \gg E_C $. Therefore, $\epsilon= \sqrt{8E_C/E_J} \ll 1 $, and one can express $\cos(\hat{\phi}/\phi_0)$ in terms of normal ordered operator products as

\begin{equation}
\cos\left(\sqrt{\frac{\epsilon}{2}} (\hat{a}+\hat{a}^\dag) \right)= e^{-\frac{\epsilon}{4}}  \sum_{n,m; n+m={\rm even}}^\infty   \frac{\left(-\frac{\epsilon}{2} \right)^{\frac{n+m}{2}}}{n! m!}     \left(\hat{a}^{\dag}\right)^n \hat{a}^m,   
\end{equation} 

If one keeps the number conserving operators only, the nonlinear part of Hamiltonian gets the following form as

\begin{equation}
\hat{\mathcal{H}}_1\simeq \hbar \delta \omega_0 \hat{a}^\dag \hat{a} -\hbar\Omega  \hat{a}^\dag \hat{a}^\dag  \hat{a} \hat{a} +  \frac{\hbar\Omega'}{6} \hat{a}^\dag \hat{a}^\dag \hat{a}^\dag \hat{a} \hat{a} \hat{a} + \dots,
\end{equation} 

In this last equation the frequency shift and interaction energies are given by 

\begin{equation}
\delta \omega_0=\sqrt{2E_J E_C}\left(1-e^{-\frac{\epsilon}{4}}\right),\qquad \Omega=\frac{E_C}{ 2} e^{-\frac{\epsilon}{4}},\qquad \Omega'=  \frac{\epsilon}{3} \Omega.
\end{equation} 

The harmonic frequency shift can be absorbed into a redefinition of $\omega_0$, i.e. $\omega_0+\delta\omega_0-\Omega\rightarrow \omega_0$, and for low excitation numbers the transmon Hamiltonian would be simplified as

\begin{equation}
\hat{\mathcal{H}}_{\rm NLC}\simeq \hbar\omega_0 \hat{a}^\dag a  - \hbar\Omega \left( \hat{a}^\dag  \hat{a} \right)^{2}.
\end{equation} 

\subsection{Coupled Circuit}
Consider a circuit of two transmon quibts coupled together via a capacitor and an inductor. To distinguish the variables we use $(\phi_{l},q_l)$ and $(\phi_r,q_r)$ for the flux and charge of the left and right transmon, respectively. The Lagrangian of the full circuit is given by 

\begin{equation}
\mathcal{L}= \left[ \frac{C_l}{2} \dot \phi_{l}^2 + E_{Jl} \cos \left( \frac{\phi_l}{\phi_0}\right) \right]+ \left[ \frac{C_r}{2} \dot \phi_{r}^2 + E_{Jr} \cos \left( \frac{\phi_r}{\phi_0}\right) \right]+ \frac{C_J}{2} (\dot \phi_{r}-\dot \phi_{l})^2  - \frac{1}{2L_J} \left( \phi_{r}-\phi_{l} \right)^{2},
\end{equation}
where $C_{\eta}$ and $E_{J \eta}$ denote the capacitance and Josephson junction energy of each sub-circuit $\eta=l,r$, and $L_J$ and $C_J$ are the inductance and capacitance of the coupling branch. We introduce the node charges $Q_\eta =\frac{\partial \mathcal{L}}{\partial \dot \phi_\eta}$ fulfilling commutation relations $[\hat{\phi}_\eta,\hat{Q}_{\eta^\prime}]=i\hbar \delta_{\eta,\eta'}$. By introducing a vector notation $\hat{\vec \phi} \equiv (\hat{\phi}_l,\hat{\phi}_r)$ and $\hat{ \vec Q} \equiv (\hat{Q}_l,\hat{Q}_r)$, the equivalent Hamiltonian can be written as 

\begin{equation}\label{capacitive-inductive coupling}
\hat{\mathcal{H}} =  \frac{1}{2} \hat{ \vec Q} \,\mathcal{C}^{-1} \, \hat{ \vec Q}^T - \left[  E_{Jl} \cos \left( \frac{\hat{\phi}_l}{\phi_0}\right) +  E_{Jr} \cos \left( \frac{\hat{\phi}_r}{\phi_0}\right) \right] + \frac{1}{2L_J} \left( \hat{\phi}_{r}-\hat{\phi}_{l} \right)^{2},
\end{equation}
where $C$ is the capacitance matrix given by the following equation
\begin{equation}
\mathcal{C}=
\left(\begin{array}{cc}
C_l +C_J  &   - C_J           \\
-C_J           & C_r +C_J     \\
\end{array}\right).
\end{equation}

Equation~\ref{capacitive-inductive coupling} can be further simplified to get the following Hamiltonian for two coupled transmons:

\begin{align*}
\hat{\mathcal{H}} &= \left[ \frac{1}{2}\frac{C_r + C_J}{C_rC_l + CrC_J + C_lC_J} \hat{Q}_l^2 -  E_{Jl} \cos \left( \frac{\hat{\phi}_l}{\phi_0}\right) + \frac{1}{2L_J}  \hat{\phi}_l^2  \right]  \\
&+ \left[ \frac{1}{2}\frac{C_l + C_J}{C_rC_l + CrC_J + C_lC_J} \hat{Q}_r^2 -  E_{Jr} \cos \left( \frac{\hat{\phi}_r}{\phi_0}\right) + \frac{1}{2L_J}  \hat{\phi}_r^2 \right] \\
&+ 
\frac{C_J}{C_rC_l + CrC_J + C_lC_J}\hat{Q}_r \hat{Q}_l - \frac{1}{L_J} \hat{\phi}_{r} \hat{\phi}_{l},
\end{align*}

Without inductive coupling, the first two terms in each bracket are Hamiltonians of two transmons with modified shunt capacitors, and the last term describes the interaction Hamiltonian via capacitive coupling. As can be seen both of the capacitive and conductive couplings have the same form and only the sign of the interaction is different. Therefore, without loss of generality one can consider one type of coupling only, and the results are properly applicable to the other type via duality. In what follows we focus on the inductive case, i.e. $C_J = 0$. 

The presence of coupling inductance $L_J$, modifies the effective inductance of each transmon hence, the natural frequency of each qubit would be given as

\begin{equation}
\omega_{0_{l/r}} = \sqrt{\frac{1}{C_{l/r}}\left( \frac{1}{L_J} + \frac{E_{J_{l/r}}}{\phi_0^2} \right)} = \frac{1}{\sqrt{C_{l/r} L_{l/r}^t}}
\end{equation}

Using the normalised variables as in Eq.~\ref{normalized variables} and their corresponding bosonic operators the coupled qubits dynamics is determined via the following Hamiltonian

\begin{equation}
\label{inductor coupled Hamiltonian}
\hat{\mathcal{H}} = \hbar \omega_{0l}\left(\hat{a}^\dagger_l \hat{a}_l + \frac{1}{2} \right) + \hbar \omega_{0r}\left(\hat{a}^\dagger_r\hat{a}_r + \frac{1}{2} \right) - \frac{\hbar}{2}\sqrt{\frac{L_l^t L_r^t}{L_J^2}}\sqrt{\omega_{0l} \omega_{0r}}\left(\hat{a}_r\hat{a}^\dag_l + \hat{a}_r^\dag \hat{a}_l  \right)
\end{equation}

Notice that in the last parenthesis we dropped the non-particle conserving terms of $\hat{a}_r\hat{a}_l + \hat{a}_r^\dag \hat{a}_l^\dag$, which is a valid assumption in the rotating wave approximation (RWA) limit. In this final equation the first two terms correspond to the Hamiltonian of each site on the left and right (i.e., 1$^{st}$ order approximation of the qubit) and the last term is the hopping between the coupled qubits.

Now assume that the Josephson junction energy of each transmon at each site is harmonically modulated as $E_{Jl,r}(t)=E_{Jl,r}^0 +e_{Jl,r} \cos(\omega_M t + \Phi_{0l,r}) $, leading to a harmonic modulation of the natural frequency of each qubit. Plugging this form back into Eq.~\ref{inductor coupled Hamiltonian} and assuming that $e_{J_{l,r}} \ll E_{J_{l,r}}^0$, the Hamiltonian of the two-coupled qubits would be modified as

\begin{align*}
\hat{\mathcal{H}} &= \hbar \omega_{0l}\left(1 + \frac{1}{2}\frac{e_{Jl} L_J}{\phi_0^2 + E_l^0 L_J}  \cos(\omega_M t + \Phi_{0l}) \right)\left(\hat{a}^\dagger_l \hat{a}_l + \frac{1}{2} \right)  \\
&+\hbar \omega_{0r}\left(1 + \frac{1}{2}\frac{e_{Jr} L_J}{\phi_0^2 + E_r^0 L_J}  \cos(\omega_M t + \Phi_{0r}) \right)\left(\hat{a}^\dagger_r \hat{a}_r + \frac{1}{2} \right) \\
&- \frac{\hbar}{2}\sqrt{\frac{L_l^t L_r^t}{L_J^2}}\sqrt{\omega_{0l} \omega_{0r}}\left(\hat{a}_l\hat{a}^\dagger_r + \hat{a}_r \hat{a}^\dagger_l \right)
\end{align*}

Since the modulation effect on coupling terms are of second order correction, those corrections have been ignored in the first order calculation which is the main scope of this paper.

\subsection{Floquet theorem and the unitary transformation}
As shown at the end of the last section the problem of two coupled qubits with modulated Josephson junction energy can be transformed to a more general problem of two coupled bosonic degrees of freedom when the on-site energies are harmonically modulated. The nodes are coupled together via a particle-conserving operator described with an effective hopping from one node to the other. In this section we briefly review the Floquet theorem needed for the analysis in this paper. Although we present the argument for a two-site lattice only, but the same treatment is indeed applicable to an extended lattice with many nodes. The interested reader may refer to relevant references~\cite{Lindner:2011kx,Fang:2012uq,Rechtsman:2013fk} for further information and elaboration on Floquet theorem. 

The following Hamiltonian gives the most general form of two bosonic nodes with harmonically modulated on-site energies and coupled together with a fixed coupling rate $J$.

\begin{equation}
\label{tight-binding}
\begin{split}
\hat{H}(t)/\hbar=& -J \left( \hat{a}^{\dagger}_2 \hat{a}_1 + \hat{a}^{\dagger}_1 \hat{a}_2 \right) \\
&+ \left[ \omega_{01} + \Omega_{01} \cos{\left( \omega_M t + \phi_1 \right)}\right] \left( \hat{a}^{\dagger}_1 \hat{a}_1 + \frac{1}{2}\right) 
+ \left[ \omega_{02} + \Omega_{02} \cos{\left( \omega_M t + \phi_2 \right)}\right] \left( \hat{a}^{\dagger}_2 \hat{a}_2 + \frac{1}{2}\right)
\end{split}
\end{equation}

For every node described as a harmonic oscillator, Fock space is the eigen-space of number operator satisfying the following relation

\begin{equation}\label{Fock states}
\omega_m \left( \hat{a}^{\dagger}_m \hat{a}_m + \frac{1}{2}\right) \ket{n}_m =  \omega_m \left( n_m + \frac{1}{2}\right) \ket{n}_m
\end{equation}
where the index $m$ refers to the $m^{th}$ node in the lattice. 

Now consider the situation where the characteristic frequency of each bosonic degree of freedom is harmonically modulated in time with frequency $\omega_M$, and follows the general form of $\omega_m=\omega_{0m} + \Omega_{0m} \cos{\left( \omega_M t + \phi_m\right)}$. For each harmonic oscillator we define the following rotated Fock states

\begin{equation}
\ket{N}_m = \ket{n}_m \exp{\left[-i~ \omega_{0m}~(n_m+1/2)t \right]} \exp{\left[ -i~\frac{\Omega_{0m}}{\omega_M} (n_m + 1/2) \sin{(\omega_M t + \phi_m)} \right]} 
\end{equation}

This wave-function composed of three main parts: 1) the Fock state, 2) a free-propagation of the Fock state given by the first exponential, and 3) a time-harmonically modulated rotation given by the second exponential. It is straight-forward to show that this wave-function is a solution of the periodically modulated harmonic oscillator Hamiltonian of Eq.~\ref{tight-binding}. In other words the aforementioned time-harmonic modulation of the trap frequency changes the instantaneous frequency of each Fock state and the new basis are related to the old ones via the following transformation

\begin{equation}
R_m(t) = \exp\left[-i ~ \frac{\Omega_{0m}}{\omega_M} \left( n_m + 1/2 \right) \sin{(\omega_M t + \phi_m)}\right].
\end{equation} 

In this rotated basis frame the transformed Hamiltonian reads as

\begin{equation}
\hat{H}_{\text{rot}}(t) = \hat{U}^\dag(t) \hat{H}(t) \hat{U}(t) - i\hbar \hat{U}^\dag(t) \dot{\hat{U}}(t)
\end{equation}
where $\hat{U}(t)= \otimes_{m} R_m(t)$ is the unitary transformation with $R_m(t)$ elements. The above relation combined with the tight-binding Hamiltonian of Eq.~\ref{tight-binding} leads to the following Hamiltonian in the rotated frame

\begin{equation}
\label{rotating}
\hat{H}_{\text{rot}}(t)/\hbar = -J  \exp{\left(i~  \left[\frac{\Omega_{02}}{\omega_M} \sin{\left( \omega_M t + \phi_2 \right)} -  \frac{\Omega_{01}}{\omega_M} \sin{\left( \omega_M t + \phi_1 \right)} \right]\right) } \hat{a}^{\dagger}_2 \hat{a}_1 + \text{h.c.} 
\end{equation}
where the fixed phase difference of $e^{i \left(\omega_{02} - \omega_{01} \right) t }$ has been dropped in the last equation for sake of simplicity. In other words the modulation of the on-site energy of each node in a lattice can be translated to an effective modulation of the hopping rate between nearest neighbours.

Similar to the original Hamiltonian in Eq.~\ref{tight-binding}, the rotated Hamiltonian in Eq.~\ref{rotating} is also periodic in time, i.e., $\hat{H}_{\text{rot}}\left( t+\frac{2\pi K}{\omega_M} \right)=\hat{H}_{\text{rot}}(t)$. Therefore, the solutions are pseudo-periodic functions in time having the general form of $\ket{\psi(t)}=\ket{\psi(t)}_p e^{-i\epsilon_p t/\hbar}$, where $\ket{\psi(t)}_p$ is a periodic function in time with the same periodicity of the Hamiltonian and $\epsilon_p$ is the quasi-energy. Moreover, for any pair of the quasi-energy and the eigenfunction as $(\epsilon_p,\ket{\psi(t)}_p)$ there are infinitely many solution pairs for any integer $K$ satisfying the following form

\begin{equation}
\left(\epsilon_p + \hbar K\omega_M ,~ \ket{\psi(t)}_p e^{iK\omega_M t} \right) 
\end{equation}

In other words, for every energy state $\epsilon_p$ within the irreducible Brillouin zone there are infinitely many other levels in the other zones separated from each other by $\hbar K \omega_M$. The corresponding wave-functions of these states are related to the wave-function of the main zone via a $e^{iM\omega_M t}$ phase factor, hence remaining periodic in time as expected. 

For the sake of simplicity, let's assume that the nodes are identical (i.e., $\omega_{01}=\omega_{02}$) and the modulation depth is the same for all nodes (i.e., $\Omega_{02}=\Omega_{01} = \Omega_0$). Knowing that $e^{ix\sin{\theta}} = \sum_{n=-\infty}^{\infty} \mathcal{J}_{n}\left( x \right) e^{in\theta}$, where $\mathcal{J}_{n}\left( x \right)$ is the $n^{th}$-order Bessel function of first kind, the periodic Hamiltonian in Eq.~\ref{rotating} can be expressed in terms of the stationary partial Hamiltonians with effective coupling $J_n$ after averaging over one time-period as

\begin{equation}
\begin{split}
J_n = J \frac{\omega_M}{2\pi} \int^{\frac{2\pi}{\omega_M} + \tau}_{\tau} dt ~ e^{i n \omega_M t} \exp{\left[i\left(\frac{2\Omega_0}{\omega_M} \sin(\frac{\phi_2-\phi_1}{2}) \sin(\omega_M t + \frac{\pi + \phi_1+\phi_2}{2}) \right)\right]} \\
= i^n e^{i n \frac{\phi_1+\phi_2}{2}} \mathcal{J}_n\left(\frac{2\Omega_0}{\omega_M} \sin(\frac{\phi_2-\phi_1}{2}) \right) 
\end{split}
\end{equation}

Due to the monotonic decrease of the effective coupling rates $J_n$, the study can be limited to the lowest order partial Hamiltonian, simplifying the Eq.~\ref{rotating} to an effective Hamiltonian as

\begin{equation}
\label{effective}
\hat{H}/\hbar= i~ J e^{i \frac{\phi_1+\phi_2}{2}} \mathcal{J}_{1}\left(\frac{2\Omega_0}{\omega_M} \sin(\frac{\phi_2-\phi_1}{2}) \right) \hat{a}^{\dagger}_2 \hat{a}_1 + \text{h.c.} 
\end{equation}

This derivation shows how the on-site energy modulation can be translated to a non-trivial change in the tunnelling properties of the lattice. Specifically the non-vanishing phase of the effective coupling indicates that the coupling rate is asymmetric and $i^{th} \rightarrow \j^{th}$ coupling rate is not the same as for the reverse direction. 

\section{Target model: The bosonic Creutz ladder}

The possibility to manipulate the phase of the coupling term in a network of superconducting qubits allows ones to simulate fundamental problems in high energy physics as well as the condensed matter. For instance, quantum Hall effect, topological insulators, and chiral edge modes are some of the important phenomena that could be investigated in these circuits. Using the general machinery explained and developed in the previous parts, in this section we are focusing on a particular building block of a model that shows this multidisciplinary physics: the bosonic Creutz ladder~\cite{PhysRevLett.83.2636,PhysRevLett.102.135702,PhysRevX.7.031057}. This model describes a cross-linked ladder in a classical magnetic field. Due to its structure and the interference effects, isolated edge states can appear depending on the values of the hopping and the magnetic field. In fact, there is a deep connection between the domain-wall approach of the chiral modes in lattice gauge theory, and the robust nature of these states under small variations of the bond strengths; this feature is linked to the topological properties of the ladder.

This model is defined in a two-leg ladder, with a Hamiltonian given by

\begin{equation}\label{Creutz model}
H= - \sum_{n} \left[ t_{d} \left( b^{\dagger}_{n} a_{n+1} + a^{\dagger}_{n} b_{n+1} \right) +  e^{i \phi} \left(a^{\dagger}_{n+1} a_{n} + b^{\dagger}_{n} b_{n+1} \right) + \frac{t_{v}}{2} \left( b^{\dagger}_{n} a_{n} + a^{\dagger}_{n+1} b_{n+1}\right) + \text{h.c.} \right], 
\end{equation}

\begin{figure}
\includegraphics[scale=1]{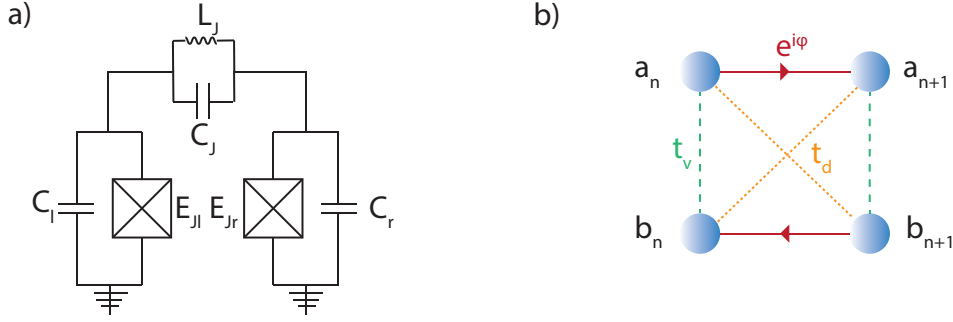}
\caption{\label{fig1} (a) a circuit of two transmons coupled with both inductive and capacitive elements. (b) Schematics of a plaquette from a bosonic Creutz ladder showing all the nodes and the coupling terms.}
\end{figure}

As schematically shown in Fig.~\ref{fig1}(b), $a_{n}$ and $b_{n}$ are the bosonic degrees of freedom in a two-leg ladder with $t_{d}$ and $t_{v}$ being the hopping terms in the diagonal and vertical directions, respectively. Moreover, $\phi$ is the magnetic flux. In the Fourier basis the Hamiltonian of an $N$-site ladder could be written as

\begin{equation}\label{Fourier description}
H=  \sum_{k} \begin{pmatrix} a^{\dagger}_{k}, & b^{\dagger}_{k} \end{pmatrix}
\begin{pmatrix} 
-2 \cos{\left( \frac{2\pi k}{N} + \phi \right)}  &  -2 t_{d} \cos{\left( \frac{2\pi k}{N} \right)} - t_{v} \\
-2 t_{d} \cos{\left( \frac{2\pi k}{N} \right)} - t_{v} & -2 \cos{\left( \frac{2\pi k}{N} - \phi \right)}
\end{pmatrix}
\begin{pmatrix} a_{k} \\ b_{k} \end{pmatrix} =  \sum_{k} \begin{pmatrix} a^{\dagger}_{k}, & b^{\dagger}_{k} \end{pmatrix} \vec{n}_{k} \vec{\sigma} \begin{pmatrix} a_{k} \\ b_{k} \end{pmatrix},
\end{equation}
where $\sigma^{(\alpha)}$ is the $\alpha$-Pauli matrix, $n^{(0)}_{k}=-2 \cos{\left( \frac{2\pi k}{N} \right)} \cos{\left( \phi \right)} $, $n^{(x)}_{k}=-2 t_{d} \cos{\left( \frac{2\pi k}{N} \right)} - t_{v}$, and $n^{(z)}_{k}=2 \sin{\left( \frac{2\pi k}{N} \right)} \sin{\left( \phi \right)}$. 

After diagonalization, the Hamiltonian reads as follows

\begin{equation}
H=  \sum_{k} \begin{pmatrix} \eta^{\dagger}_{+,k} & \eta^{\dagger}_{-,k} \end{pmatrix}
\begin{pmatrix} 
+\Lambda_{k} - 2 \cos{\left( \frac{2\pi k}{N} \right)} \cos{\left( \phi \right)}  & 0 \\
0 & -\Lambda_{k} - 2 \cos{\left( \frac{2\pi k}{N} \right)} \cos{\left( \phi \right)} 
\end{pmatrix}
\begin{pmatrix} \eta_{+,k} \\ \eta_{-,k} \end{pmatrix},
\end{equation}
where $\Lambda_{k} =\sqrt{4 \sin^{2}{\left( \frac{2\pi k}{N} \right)} \sin^{2}{\left( \phi \right)} + \left[ 2 t_{d} \cos{\left( \frac{2\pi k}{N} \right)}+ t_{v} \right]^{2}} $.

Depending on the parameter values in the Hamiltonian the system acquires different symmetries. We define $h_{k} =\vec{n}_{k} \vec{\sigma}$. From the Fourier description of Eq.~\ref{Fourier description}, it is clear that $\sigma^{x} h_{k} \sigma^{x} = h_{-x}$, corresponding to the time reversal symmetry for any parameter values of the Hamiltonian. 

At $\phi = \frac{\pi}{2}$, one can obtain two additional symmetries as $\sigma^{z} h_{k} \sigma^{z} = - h_{-x}$ and $\sigma^{y} h_{k} \sigma_{y}= -h_{k}$ corresponding to the particle-hole symmetry and chiral symmetry, respectively. At this value $n^{(x)}_{k}=-2 t_{d} \cos{\left( \frac{2\pi k}{N} \right)} - t_{v}$ and $n^{(z)}_{k}=2 \sin{\left( \frac{2\pi k}{N} \right)}$ are the non-zero values of the Hamiltonian in Eq.~\ref{Fourier description}. 

The chiral symmetry implies that any eigenstate $|E\rangle$ with energy $E$ has a counterpart $|-E\rangle = \sigma^{y} |E\rangle$ with energy $-E$. Therefore, in a chiral-symmetric system the eigenstates come in pairs at $\pm E$. For a state at $E=0$, the state is its own partner, i.e. $|0\rangle = \sigma^{y} |0\rangle$. To explicitly construct this zero mode, we will use the low energy continuum theory. We consider the limit $|t_{v}| < |t_{d}| < 1$ and focus on the low energy states near $\frac{2\pi k}{N\alpha}=\frac{\pi}{\alpha} + q$, with lattice spacing $\alpha$ and small $q$. In real space $q \to -i \partial_{x}$ and $H \to - i v_{F} \sigma^{z} \partial_{x} + m \sigma^{x}$, with $v_{F}=2\alpha$ and $m=2 t_{d} - t_{v}$. To describe the zero mode we allow $m\left( x \right)$ to vary spatially with a kink such that $m\left( x \to + \infty \right) > 0$ and $m\left( x \to - \infty \right) < 0$. A zero energy solution $H|0\rangle =0$ can be constructed considering eigenstates $|\pm y \rangle$ of $\sigma^{y}$ with eigenvalue $\pm 1$, giving $\partial_{x} \psi_{0,\pm} \left( x \right) =\pm \frac{m\left( x \right)}{v_{F}} \psi_{0,\pm} \left( x \right)$. Integrating the first-order equation leads to a single normalisable solution as $\psi_{0,-} \left( x \right)= e^{- \int_{0}^{x} dx' \,  \frac{m\left( x' \right)}{v_{F}}}$. This solution is a a localised wave-function at $x=0$ with exponentially decaying tails on the sides.

Due to the periodicity of the Hamiltonian in $k$, all integer $k$ within $(-N/2,+N/2]$ define a closed curve in the $(n^{(x)}_{k}-n^{(z)}_{k})$-plane whose features strongly depends on $ \mathcal{R} = t_{v} /t_{d}$ ratio. If $ \left| \mathcal{R} \right| <2 $, this curve will enclose the origin $(0,0)$  where the Hamiltonian is strictly zero. On the other hand, if $ \left| \mathcal{R} \right| >2$ the curve will not enclose this point and is deformed to a trivial one. When $ \left| \mathcal{R} \right| <2 $, the number of times the closed curve winds around the origin defines a topological invariant, the \emph{``winding''} number $\nu$.

Another easy way of characterising the topological properties of the Creutz ladder is taking the ``strong'' coupling limit in the lattice description, when $\phi = \frac{\pi}{2}$, $t_{v}=0$, and $t_{d}=1$. In this limit the Hamiltonian of Eq.~\ref{Fourier description} recasts into

\begin{equation}
\begin{split}
H =& - \sum_{n} \left( b_{n}^{\dagger} a_{n+1} + a_{n}^{\dagger} b_{n+1} + a_{n+1}^{\dagger} b_{n} + b_{n+1}^{\dagger} a_{n} \right) - i \sum_{n} \left( a_{n}^{\dagger} a_{n+1} + b_{n+1}^{\dagger} b_{n} - a_{n+1}^{\dagger} a_{n} - b_{n}^{\dagger} b_{n+1} \right) \\
=& \sum_{k} \begin{pmatrix} a_{k}^{\dagger}, & b_{k}^{\dagger} \end{pmatrix}
\begin{pmatrix}
2 \sin{\left( \frac{2 \pi k}{N} \right)} & -2 \cos{\left( \frac{2 \pi k}{N} \right)} \\
-2 \cos{\left( \frac{2 \pi k}{N} \right)} & - 2 \sin{\left( \frac{2 \pi k}{N} \right)}
\end{pmatrix}
\begin{pmatrix} a_{k}  \\ b_{k}  \end{pmatrix} \\
=& \sum_{k} \begin{pmatrix} \eta_{+,k}^{\dagger}, & \eta_{-,k}^{\dagger} \end{pmatrix}
\begin{pmatrix}
2  & 0 \\
0 & - 2 
\end{pmatrix}
\begin{pmatrix} \eta_{+,k}  \\ \eta_{-,k}  \end{pmatrix} 
=\sum_{n} \begin{pmatrix} \eta_{+,n+1/2}^{\dagger}, & \eta_{-,n+1/2}^{\dagger} \end{pmatrix}
\begin{pmatrix}
2  & 0 \\
0 & - 2 
\end{pmatrix}
\begin{pmatrix} \eta_{+,n+1/2}  \\ \eta_{-,n+1/2}  \end{pmatrix} 
\end{split}
\end{equation}
with ``Bloch'' basis as

\begin{equation}
\begin{pmatrix} \eta_{+,k}  \\ \eta_{-,k}  \end{pmatrix} 
= \begin{pmatrix}
 \cos{\left( \frac{ \pi k}{N} - \frac{\pi}{4} \right)} &  \sin{\left( \frac{\pi k}{N} - \frac{\pi}{4} \right)} \\
 \sin{\left( \frac{ \pi k}{N} - \frac{\pi}{4} \right)} & -  \cos{\left( \frac{ \pi k}{N} - \frac{\pi}{4}\right)}
\end{pmatrix}
\begin{pmatrix} a_{k}  \\ b_{k}  \end{pmatrix}
\end{equation}
and the maximally localised ``Wannier'' basis in the following form

\begin{equation}
\begin{split}
\begin{pmatrix} \eta_{+,n+1/2}  \\ \eta_{-,n+1/2}  \end{pmatrix} 
=&\frac{1}{\sqrt{N}} \sum_{k} e^{i2\pi k \left( n+1/2 \right)/N} 
\begin{pmatrix}
 \cos{\left( \frac{ \pi k}{N} - \frac{\pi}{4} \right)} &  \sin{\left( \frac{\pi k}{N} - \frac{\pi}{4} \right)} \\
 \sin{\left( \frac{ \pi k}{N} - \frac{\pi}{4} \right)} & -  \cos{\left( \frac{ \pi k}{N} - \frac{\pi}{4}\right)}
\end{pmatrix}
\begin{pmatrix} a_{k}  \\ b_{k}  \end{pmatrix}\\
=& \frac{1}{2} \begin{pmatrix}
 e^{-i \pi /4} \left( a_{n+1} - b_{n} \right) + e^{i \pi /4} \left( a_{n} - b_{n+1} \right)  \\
-e^{i \pi /4} \left( a_{n+1} + b_{n} \right) - e^{-i \pi /4} \left( a_{n} + b_{n+1} \right)  
\end{pmatrix}
\end{split}
\end{equation}

Since the Wannier functions are the Fourier transforms of the Bloch functions, one can show that the center of the maximally localised Wannier function gives the Berry phase of the band according to the following relation
 
\begin{equation}
\begin{split}
| \eta_{-,n+1/2} \rangle =& \frac{1}{\sqrt{N}} \sum_{k} e^{i2\pi k \left( n+1/2 \right)/N} |  \eta_{-,k} \rangle,\\
\langle  \eta_{-,n+1/2} | m |  \eta_{-,n+1/2} \rangle =& -\frac{i}{2\pi} \sum_{k} \langle \eta_{-,k} | \frac{\partial}{\partial k}  | \eta_{-,k} \rangle = \frac{\phi_{\text{Berry}}}{2 \pi}
\end{split}
\end{equation}

Explicitly in the lattice we have

\begin{equation}
\begin{split}
&\langle  \eta_{-,n+1/2} | m |  \eta_{-,n+1/2} \rangle = \\
=&\left[ e^{-i \pi /4} \left( \langle a_{n+1} | + \langle b_{n}| \right) + e^{i \pi /4} \left( \langle a_{n}| + \langle b_{n+1}| \right) \right]
\frac{m}{4}
\left[ e^{i \pi /4} \left(| a_{n+1} \rangle + | b_{n} \rangle \right) + e^{-i \pi /4} \left( | a_{n} \rangle +| b_{n+1} \rangle \right) \right]\\
=&\frac{1}{4}  \left[ \langle a_{n+1} | (n+1) | a_{n+1} \rangle +\langle b_{n}| n  | b_{n} \rangle +   \langle a_{n}| n | a_{n} \rangle + \langle b_{n+1}| (n+1) | b_{n+1} \rangle \right] = n + \frac{1}{2}
\end{split}
\end{equation}

Therefore, the Berry phase of the lower band in the Creutz ladder is given by $\phi_{\text{Berry}} - 2 \pi n  = \pi$.

From the Wannier operators we can see that in the absence of the coupling between the first and the last site of the ladder, there are two zero modes described via the following relations

\begin{equation}
\eta_{0,L} = \frac{1}{\sqrt{2}} \left( e^{i \pi /4} a_{1} + e^{-i \pi/4} b_{1} \right); ~ \, ~ \eta_{0,R} = \frac{1}{\sqrt{2}}  \left( e^{-i \pi /4} a_{N} + e^{i \pi/4} b_{N} \right).
\end{equation}

\section{Four sites building block}

In the previous section we introduced one of the important Hamiltonians of great importance in different areas of physics. Later we summarised some of the unique properties of such a ladder in supporting the chiral edge modes. In This section we limit our study to just one plaquette of the ladder as the smallest building block possessing some of the non-trivial features of the main ladder.

For a 4-site building block, we can use another setup based on a SQUID which parametrically couples the total flux in a cavity~\cite{Wilson:2011fk,PhysRevLett.109.183901}, $\hat{\Phi}_{c}$, to the pump flux, $\hat{\Phi}_{p}$ through its Hamiltonian as

\begin{equation}
\hat{H}_{SQ}=E_{J} \left| \cos{\left( \pi \hat{\Phi}_{p} / \Phi_{0} \right)} \right| \cos{\left(2 \pi \hat{\Phi}_{c} / \Phi_{0} \right)}
\end{equation}

The interaction Hamiltonian can be derived by expanding $\hat{H}_{SQ}$ to the first order in $\hat{\Phi}_{p}$ around a flux bias $\Phi_{ext}$, and to the second order in $\hat{\Phi}_{c}$ around zero. After applying the parametric approximation to the pump we get the following interaction Hamiltonian

\begin{equation}
\label{four-site}
\hat{H}_{int} = \hbar g_{0} \left( \alpha_{p} + \alpha^{*}_{p} \right) \left[\sum_{i=1}^{4} \left( \hat{a}_{i} + \hat{a}^{\dagger}_{i} \right) \right]^{2}
\end{equation}
where $\alpha_{p}$ denotes the coherent pump amplitude, the bosonic operators $\hat{a}_{i}$, $\hat{a}^{\dagger}_{i}$ is the annihilation and creation operator correspond to the four cavity modes considered here, and $g_{0}$ is an effective coupling constant. Eq.~\ref{four-site} contains a large number of terms corresponding to different physical processes. However, we can selectively activate different processes by the appropriate choice of pump frequency. If we choose to pump at the difference frequency $f_{p}=|f_{i}-f_{j}|$, $\hat{H}_{int}$ reduces to $\hat{H}_{CC} = \hbar g' \left( \hat{a}_{i} \hat{a}^{\dagger}_{j} + \hat{a}_{j} \hat{a}^{\dagger}_{i}\right) $. $\hat{H}_{CC}$ produces a coherent coupling between modes where different phases could be achieved in the Hamiltonian by phase-locking the different pump fluxes to a reference value.

From the purpose bosonic Creutz ladder, we could study the single plaquette Hamiltonian describing the dynamics of the four coupled cavity modes representing the bosonic degrees of freedom as the nodes in Fig.~\ref{fig1}(b) 

\begin{equation}
H=-\left( a^{\dagger}_{3} a_{2} + a^{\dagger}_{1} a_{4} + a^{\dagger}_{2} a_{3} + a^{\dagger}_{4} a_{1} \right)-i \left( a^{\dagger}_{1} a_{2} - a^{\dagger}_{2} a_{1} + a^{\dagger}_{4} a_{3} - a^{\dagger}_{3} a_{4} \right) = 2 \left( \eta^{\dagger}_{+} \eta_{+}  - \eta^{\dagger}_{-} \eta_{-} \right),
\end{equation}
with

\begin{equation}
\eta_{+} = \frac{1}{2} \left[ e^{-i\pi/4} \left( a_{2} - a_{3} \right) + e^{i\pi/4} \left( a_{1} - a_{4} \right) \right], ~ \, ~ \eta_{-} = \frac{1}{2} \left[ e^{i\pi/4} \left( a_{2} + a_{3} \right) + e^{-i\pi/4} \left( a_{1} + a_{4} \right) \right]
\end{equation}
and the zero modes as

\begin{equation}
\eta_{1} =\frac{1}{\sqrt{2}} \left( e^{i\pi/4} a_{1} + e^{-i\pi/4} a_{3} \right) , ~ \, ~ \eta_{2} =\frac{1}{\sqrt{2}} \left( e^{-i\pi/4} a_{2} + e^{i\pi/4} a_{4} \right).
\end{equation}

Having the eigen-energies of the Hamiltonian, it is straightforward to show that the single-particle states evolve as following 

\begin{equation}
\begin{split}
|+\rangle = \eta^{\dagger}_{+} |vac\rangle, ~ \, ~ |-\rangle = \eta^{\dagger}_{-} |vac\rangle, ~ \, ~|1\rangle = \eta^{\dagger}_{1} |vac\rangle, ~ \, ~|2\rangle = \eta^{\dagger}_{2} |vac\rangle\\
|+ (t) \rangle = e^{-i2t} |+\rangle, ~ \, ~ |- (t)\rangle = e^{i2t} |-\rangle, ~ \, ~ |1 (t)\rangle = |1\rangle, ~ \, ~ |2 (t)\rangle = |2\rangle
\end{split}
\end{equation}

Starting with the state $|a_{1}\rangle = a^{\dagger}_{1} | vac \rangle=\left[\frac{e^{i\pi/4}}{2} \eta^{\dagger}_{+} + \frac{e^{-i\pi/4}}{2} \eta^{\dagger}_{-} + \frac{e^{i\pi/4}}{\sqrt{2}} \eta^{\dagger}_{1} \right] | vac \rangle$

\begin{equation}
\begin{split}
|a_{1}(t) \rangle&= \frac{e^{i\pi/4}}{2} e^{-i2t} |+\rangle + \frac{e^{-i\pi/4}}{2}e^{i2t} |-\rangle + \frac{e^{i\pi/4}}{\sqrt{2}} |1\rangle\\
&= \frac{1}{2} \left\{  \left[1+\cos{\left(2t\right)} \right] |a_{1}\rangle + i  \left[1-\cos{\left(2t\right)} \right] |a_{3}\rangle + \sin{\left(2t\right)} \left( |a_{2}\rangle + i |a_{4}\rangle \right) \right\}
\end{split}
\end{equation}

\begin{figure}
\includegraphics[scale=0.9]{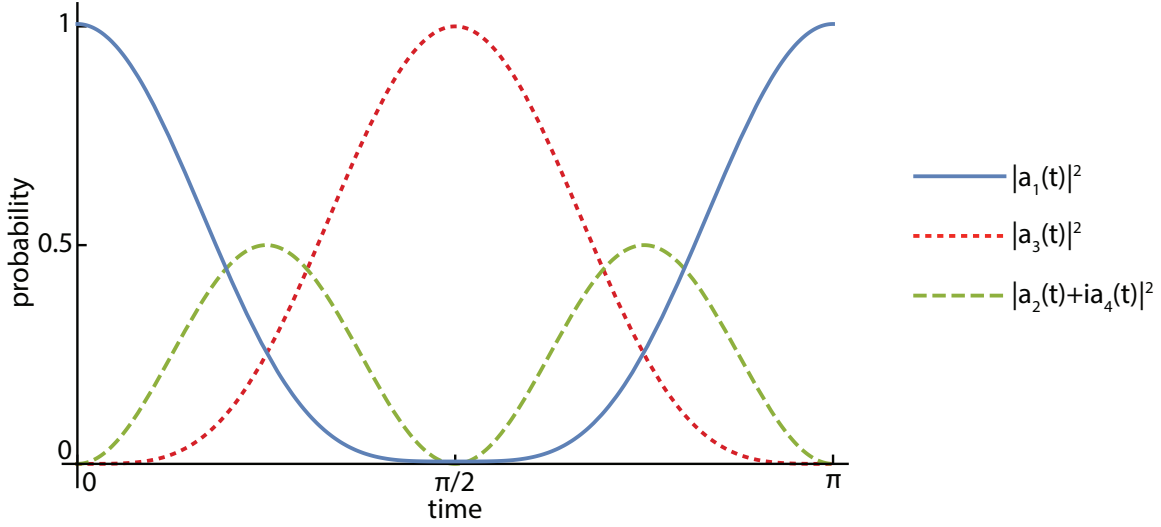}
\caption{\label{fig2} Temporal dependence of the local occupations of the modes involved in a plaquette configuration. This behaviour has some sense of chirality. Starting with an initial state $|a_{1}\rangle$, the population of this state decreases while increases $|a_{2}\rangle + i |a_{4}\rangle$ and finally the whole population appears in $|a_{3}\rangle$.}
\end{figure}

As can be seen the behaviour of the local occupations of the four modes has some signatures of the chirality. Starting at the state $|a_{1}\rangle$ at initial time, the population of this state decreases and gets transferred to the state of $|a_{2}\rangle + i |a_{4}\rangle$. Finally the whole population appears in $|a_{3}\rangle$. Figure~\ref{fig2} shows the population transfer between the states as a function of time. In other  words the time evolution of the population has some direction (clock-wise in this case) which leads to complete population transfer from $|a_{1}\rangle$ to $|a_{3}\rangle$, deterministically.  

\section{Conclusion}
In this work we proposed the possibility of realising topological features in a circuit QED setup. To imprint the chirality in a lattice of coupled transmons, we proposed a periodic modulation of the qubit fluxes at each node and showed that it leads to a complex hopping term between the adjacent sites in a lattice.

After that we introduced the bosonic Creutz ladder as an important Hamiltonian arising in several cases ranging from high-energy physics to the condensed-matter. The important ingredient of this Hamiltonian, the complex coupling, could be realised using the developed scheme in the previous section. 

Finally we investigated the simple one-plaquette, 4-site lattice of the bosonic  Creutz ladder and showed the emergence of the chiral population transfer between the bosonic degress of freedom. The proposed scheme can be implemented in the state-of-the-art results in the circuit QED. 

The scheme presented here can be simply extended to a 2D lattice where the hopping term between the adjacent sites can be easily manipulated. Moreover, by expanding the Hamiltonian of the transmon to the non-linear term one can have a proper Bose-Hubbard model, where on-site interaction exists.

As has been shown in the previous studies and in the context of highly interacting Bose-Hubbard model the effect of on-site modulation is replacing the coupling coefficient $J$  with a new, effective coupling $J_{eff}$. Since the ratio of the on-site interaction energy and the coupling $U/J_{eff}$ can tune the behaviour of the lattice in phase space all the way from a Mott insulator (for large ratio) to the super-fluid phase (for small ratio) then by changing the on-site modulation one should be able to scan over the behaviour of the system in the phase space and change the behaviour in different phases.

\section*{Acknowledgment}
H.A. acknowledges the financial support from Alexander von Humboldt foundation in terms of a postdoctoral fellowship. E.R. and E.S. acknowledge funding from MINECO/FEDER  FIS2015-69983-P and Basque  Government  IT986-16, CMW,  CWSC, and MVM acknowledge  NSERC of Canada, the Canadian Foundation for Innovation, the Ontario  Ministry of Research and Innovation, Canada First Research Excellence  Fund  (CFREF),  Industry  Canada, and the CMC for financial  support. 

\bibliographystyle{ieeetr}
\bibliography{ref}
\end{document}